\documentclass[]{spie}  

\usepackage{amsmath,amsfonts,amssymb}
\usepackage{graphicx}
\usepackage[colorlinks=true, allcolors=blue]{hyperref}
\usepackage{subcaption}
\usepackage{caption}

\title{First laboratory demonstration of real-time multi-wavefront sensor single conjugate adaptive optics}

\author[a]{Benjamin L. Gerard}
\author[b]{Daren Dillon}
\author[c]{Sylvain Cetre}
\author[b]{Rebecca Jensen-Clem}

\affil[a]{Lawrence Livermore National Laboratory}
\affil[b]{University of California Santa Cruz}
\affil[c]{Wakea Consulting}

\authorinfo{Further author information: (Send correspondence to Benjamin L. Gerard)\\ Benjamin L. Gerard: E-mail: gerard3@llnl.gov}

\begin{document} 
\maketitle

\begin{abstract}
Exoplanet imaging has thus far enabled studies of wide-orbit ($>$10 AU)  giant planet ($>$2 Jupiter masses) formation and giant planet atmospheres, with future 30 meter-class Extremely Large Telescopes (ELTs)  needed to image and characterize terrestrial exoplanets. However, current state-of-the-art exoplanet imaging technologies placed on ELTs would still miss the contrast required for imaging Earth-mass habitable-zone exoplanets around low-mass stars by ~100x due to speckle noise--scattered starlight in the science image due to a combination of aberrations from the atmosphere after an adaptive optics (AO) correction and internal to the telescope and instrument. We have been developing a focal plane wavefront sensing technology called the Fast Atmospheric Self-coherent camera Technique (FAST) to address both of these issues; in this work we present the first results of simultaneous first and second stage AO wavefront sensing and control with a Shack Hartmann wavefront sensor (SHWFS) and FAST, respectively, using two common path deformable mirrors. We demonstrate this ``multi-WFS single conjugate AO'' real-time control at up to 200 Hz loop speeds on the Santa Cruz Extreme AO Laboratory (SEAL) testbed, showing a promising potential for both FAST and similar high-speed diffraction-limited second-stage wavefront sensing technologies to be deployed on current and future observatories, helping to remove speckle noise as the main limitation to ELT habitable exoplanet imaging.
\end{abstract}

\keywords{Adaptive Optics, Wavefront Sensing, Focal Plane Wavefront Sensing, Pupil Plane Wavefront Sensing}

\section{INTRODUCTION}
\label{sec:intro}  
\begin{figure}[!h]
    \centering
    \includegraphics[width=0.5\textwidth]{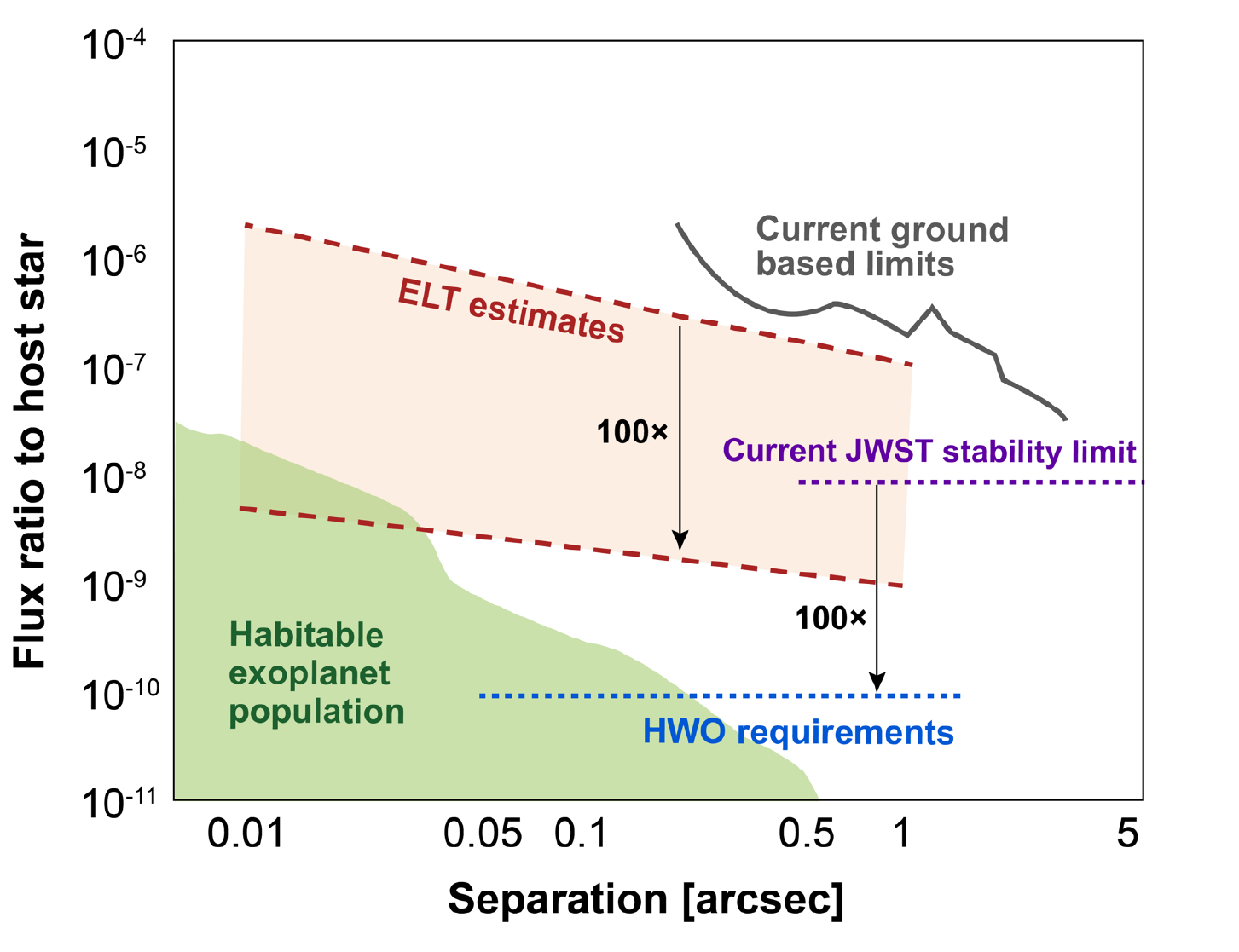}
    \caption{Adapted from Ref. \citenum{nasem2018}, exoplanet imaging detection sensitivity limits with current and future facilities---including ground-based extremely large telescopes (ELTs; Ref. \citenum{jensen-clem22}) and the spaced-based \textit{James Webb Space Telescope} (\textit{JWST}; Ref. \citenum{jwst}) and Habitable Worlds Observatory (HWO)---compared to the expected habitable exoplanet population, illustrating the 100$\times$ sensitivity gain needed for ground- or space-based habitable exoplanet detection and characterization.}
    \label{fig: intro_habfig}
\end{figure}
The 2020 Decadal Survey of Astronomy and Astrophysics (Ref. \citenum{nasem2021}) has listed exoplanet imaging technology maturation as a key priority in the next decade, with the goal of enabling habitable exoplanet detection and characterization. However, it has now become clear that both ground- and space-based state-of-the-art technologies miss the contrast requirements to enable such future detections by about two orders of magnitude, illustrated in Fig. \ref{fig: intro_habfig}.

This paper addresses this call for such habitable exoplanet imaging technology maturation by presenting a laboratory demonstration of a recently developed technology, focused on ground-based adaptive optics (AO), called multi-wavefront sensor (WFS) single conjugate AO (SCAO). We describe the concept and our laboratory setup in \S\ref{sec: concept_and_setup}, present our laboratory results in \S\ref{sec: results}, and conclude in \S\ref{sec: conclusions}.
\section{CONCEPT AND LABORATORY SETUP}
\label{sec: concept_and_setup}
This paper presents a laboratory demonstration of multi-WFS SCAO, first introduced in Ref. \citenum{gerard21a}. The concept enables multiple WFSs pointed at the same star to control one or more common path deformable mirrors (DMs), which is an alternative to a so-called ``cascaded AO'' system (Ref. \citenum{cascade}), which uses a non-common path DM that is controlled by a second WFS. The multi-WFS SCAO concept is illustrated below in Fig. \ref{fig: concept+intro}.
\begin{figure}[!h]
    \centering
    \includegraphics[width=0.4\textwidth]{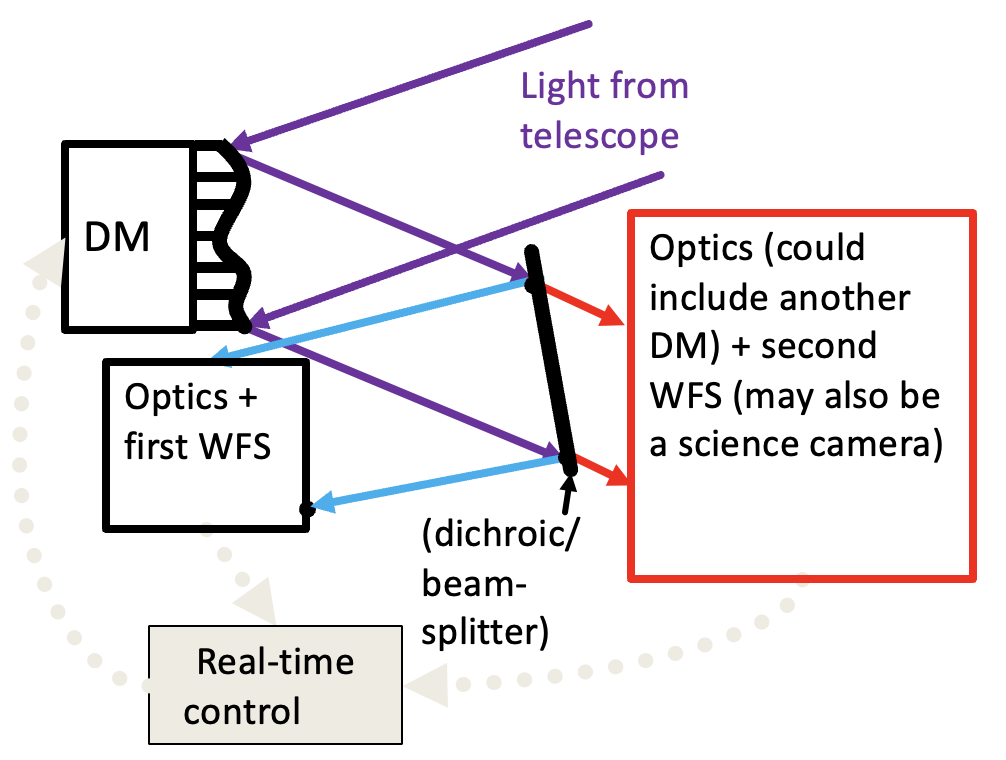}
    \caption{Illustration of Multi-WFS SCAO. Light collected from a telescope is relayed to at least one DM which is common path to at least two WFSs, which are both pointing at the same star but not necessarily operating at the same speed or wavelengths. A hard real-time control algorithm then implements spatial and temporal filtering to combine the two or more WFSs into a set of common path DM commands that prevents closed-loop instabilities from occurring.}
    \label{fig: concept+intro}
\end{figure}

\begin{figure}[!h]
    \centering
    \includegraphics[width=1.0\textwidth]{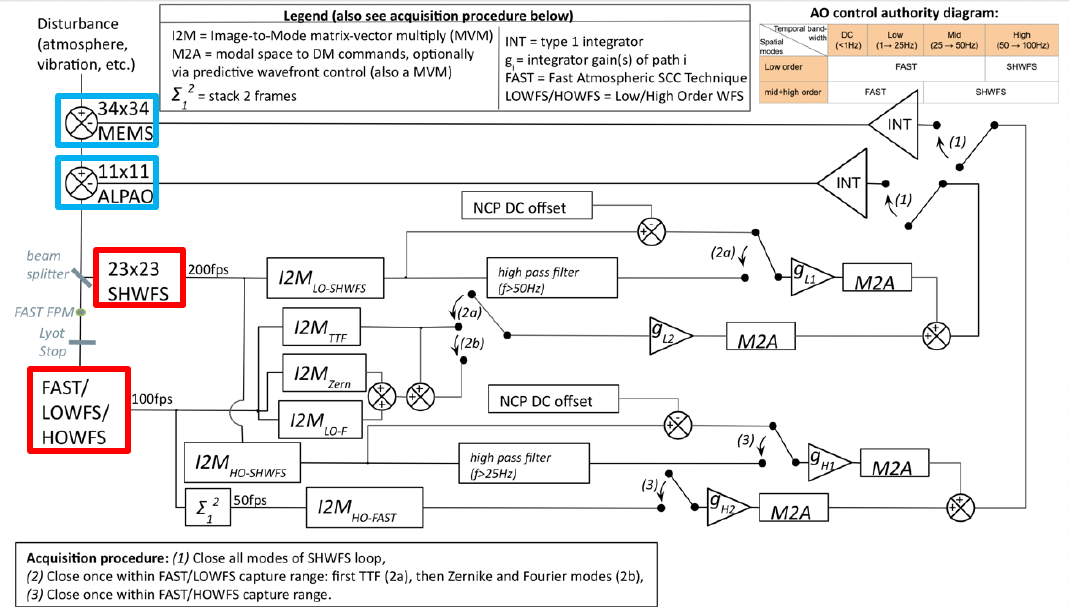}
    \caption{Multi-WFS SCAO diagram for the SEAL testbed. Two WFSs (a SHWFS and FAST, highlighted in red) control two common path DMs (highlighted in blue, although in this paper only low order spatial modes are demonstrated). Spatial and temporal filters close successively (see the acquisition procedure illustrated in the lower left) to ultimately enable a stable steady-state closed loop configuration that gives full control authority to only one WFS in a given range of spatio-temporal parameter space, illustrated by the table in the upper right of this figure.}
    \label{fig: loop_diagram}
\end{figure}

This work builds on Ref. \citenum{gerard22b}, using the identical laboratory setup on the Santa cruz Extreme AO Laboratory (SEAL; Ref. \citenum{jensen-clem21}) testbed that includes a 23$\times$23 Shack Hartmann WFS (SHWFS), a Fast Atmospheric Self-coherent camera (SCC) Technique (FAST) sensor, and for the purposes of this paper a single 11$\times$11 ALPAO DM. A 34$\times$34 MEMS DM is used in Ref. \citenum{gerard22b} for real-time high-order control for the SHWFS and FAST sensors in separate demonstrations; instead, in this paper SHWFS and FAST will be demonstrated simultaneously, although only for low order modes. More specifically, Fig. \ref{fig: loop_diagram} shows an illustration of multi-WFS SCAO applied to the SEAL testbed. 
As shown in Fig. \ref{fig: loop_diagram}, for the SEAL testbed setup used in this paper, the SHWFS runs at 200 Hz and FAST sensor runs at 100 Hz. The loops for each sensor run simultaneously in parallel Python sessions, although switches in closed-loop to different temporal filters are triggered by a clock that is pinged at the beginning of each frame in the SHWFS (i.e., the faster) loop. Spatial modes are selected as elements in a vector and must share a common modal space between WFSs. Here, for low order modes we use a modal space of tip/tilt/focus (TTF) and then a zonal space (with TTF removed) by poking ALPAO actuators for both SHWFS and FAST low order control. This is particularly convenient for FAST, since the spatial Nyquist limit for the ALPAO DM (2.5 cycles/pupil, or c/p, since the outermost ring of ALPAO actuators is blocked by the Lyot stop), is behind the 3 $\lambda/D$ radius FAST focal plane mask (FPM) diameter. 

\begin{figure}[!h]
    \centering
    \includegraphics[width=0.6\textwidth]{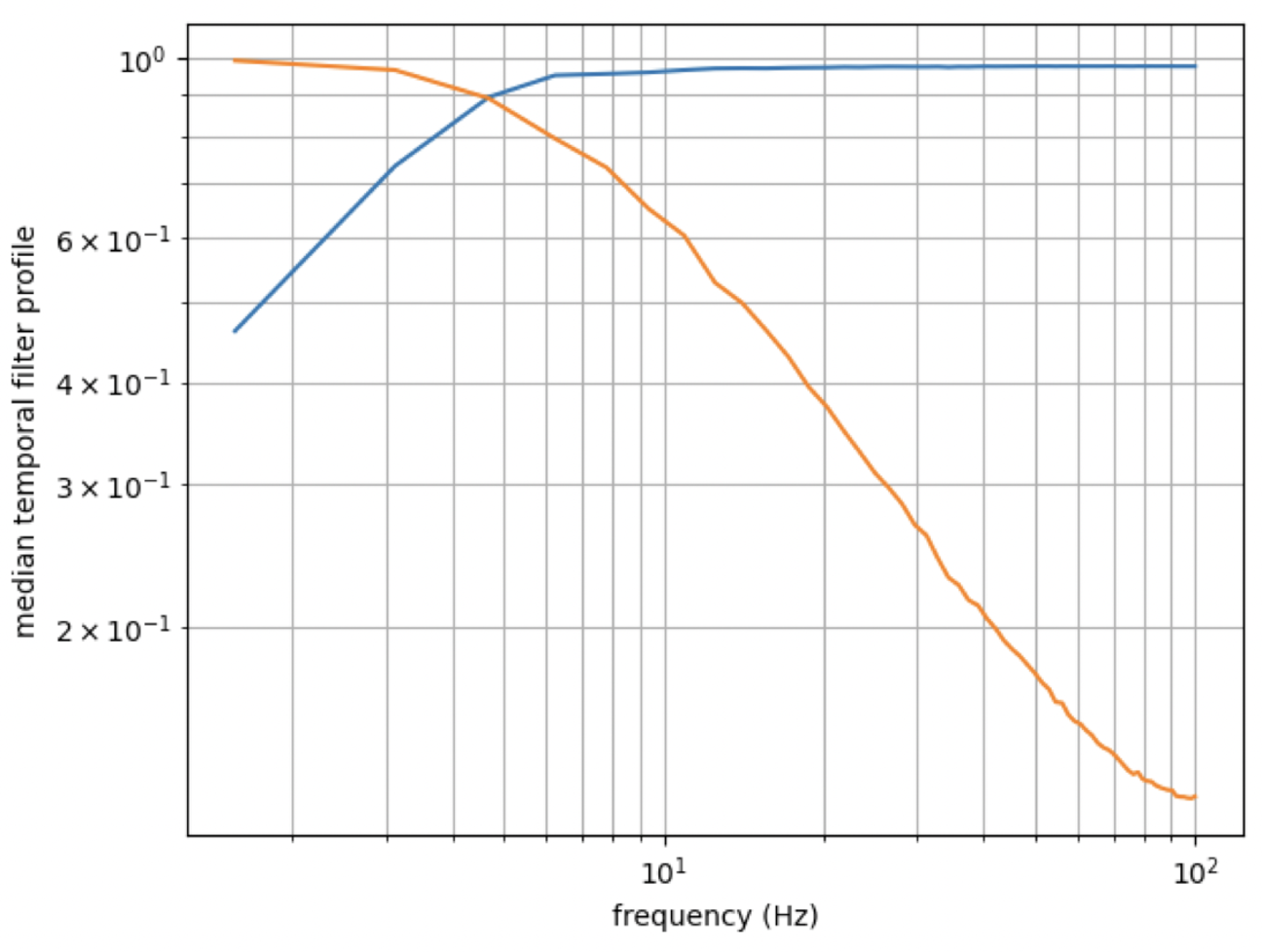}
    \caption{Low-pass (orange curve) and high-pass (blue curve) filter profiles, produced from the square root of the ratio of filtered to unfiltered power spectral density (PSD) curves. PSDs are generated using the same methods as in Ref. \citenum{gerard22b} and references therein. The two curves are medianed across all low order modal coefficients used in this paper.}
    \label{fig: bandpasses}
\end{figure}

We implement temporal filters in real-time using the same auto-regressive framework as Ref. \citenum{gavel14} that uses closed-loop coefficients generated from a matrix-vector multiplication (MVM; as described in Ref. \citenum{gerard22b}) from the current and most recent frames, $c_n$ and $c_{n-1}$, respectively, as follows:

\begin{align}
    \label{eq: filters} c_{H\{n\}} &= \alpha\; c_{H\{n-1\}} + \alpha\; ( c_n - c_{n-1})\text{, and} \\
    c_{L\{n\}} &= \alpha\; c_{L\{n-1\}} + (1-\alpha)\; c_{n-1} \text{, where}\nonumber \\
    \alpha &= e^{-\frac{f_\text{cutoff}}{f_\text{loop}}}. \nonumber
\end{align}
In equation \ref{eq: filters}, $c_{H\{n\}}$ and $c_{H\{n-1\}}$ are the respective high pass filtered coefficients at the current and previous loop iteration, and $c_{L\{n\}}$ and $c_{L\{n-1\}}$ are the equivalent for low pass filtered coefficients; $f_\text{cutoff}$ is the desired high or low pass filter cutoff frequency (at which the filter profile attenuates temporal frequencies at 1/$e$ relative to full transmission) and $f_\text{loop}$ is the loop frame rate.  Figure \ref{fig: bandpasses} shows the result of equation \ref{eq: filters} applied to our SHWFS and FAST coefficients.

Fig. \ref{fig: bandpasses} shows a filter cutoff at $\sim$6 Hz; although the filter cutoff in Fig. \ref{fig: loop_diagram} is 50 Hz, we found a lower cutoff produced better results, although this parameter space should be explored further. Fig. \ref{fig: bandpasses} also shows a discrepancy between the low and high pass filters, which do not appear to share the same 1/$e$ cutoff frequency within a factor of a few. The timing and jitter with two WFSs and two different frame rates may have contributed to this, but in this paper we did not investigate this further and found these filter profiles to enable sufficient multi-WFS SCAO performance, which we show next in \S\ref{sec: results}.
\section{RESULTS}
\label{sec: results}
\begin{figure}[!h]
    \centering
    \begin{subfigure}[b]{0.85\textwidth}
        \includegraphics[width=1.0\textwidth]{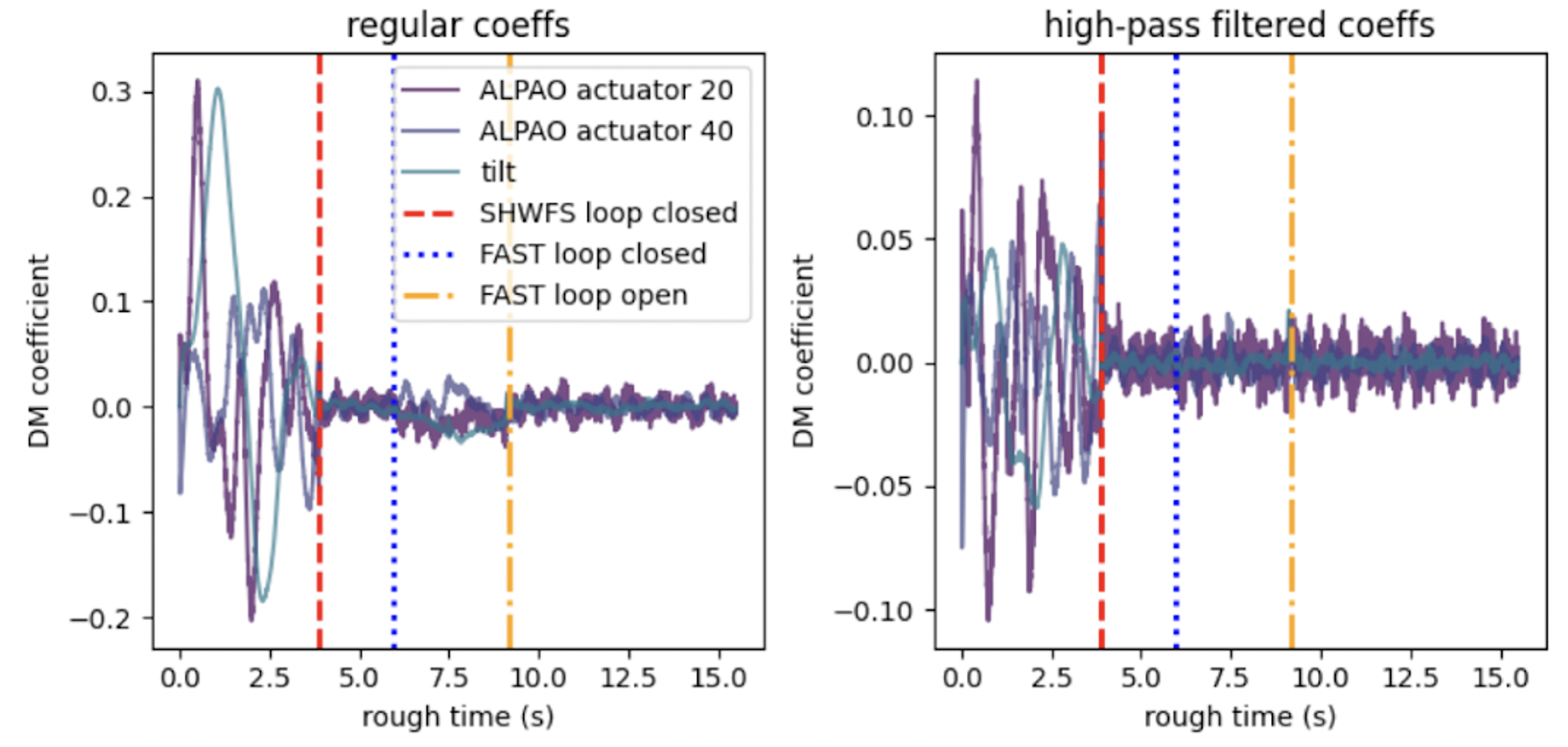}
        \caption{FAST loop gain set to zero.}
    \end{subfigure}
    \begin{subfigure}[b]{0.85\textwidth}
        \includegraphics[width=1.0\textwidth]{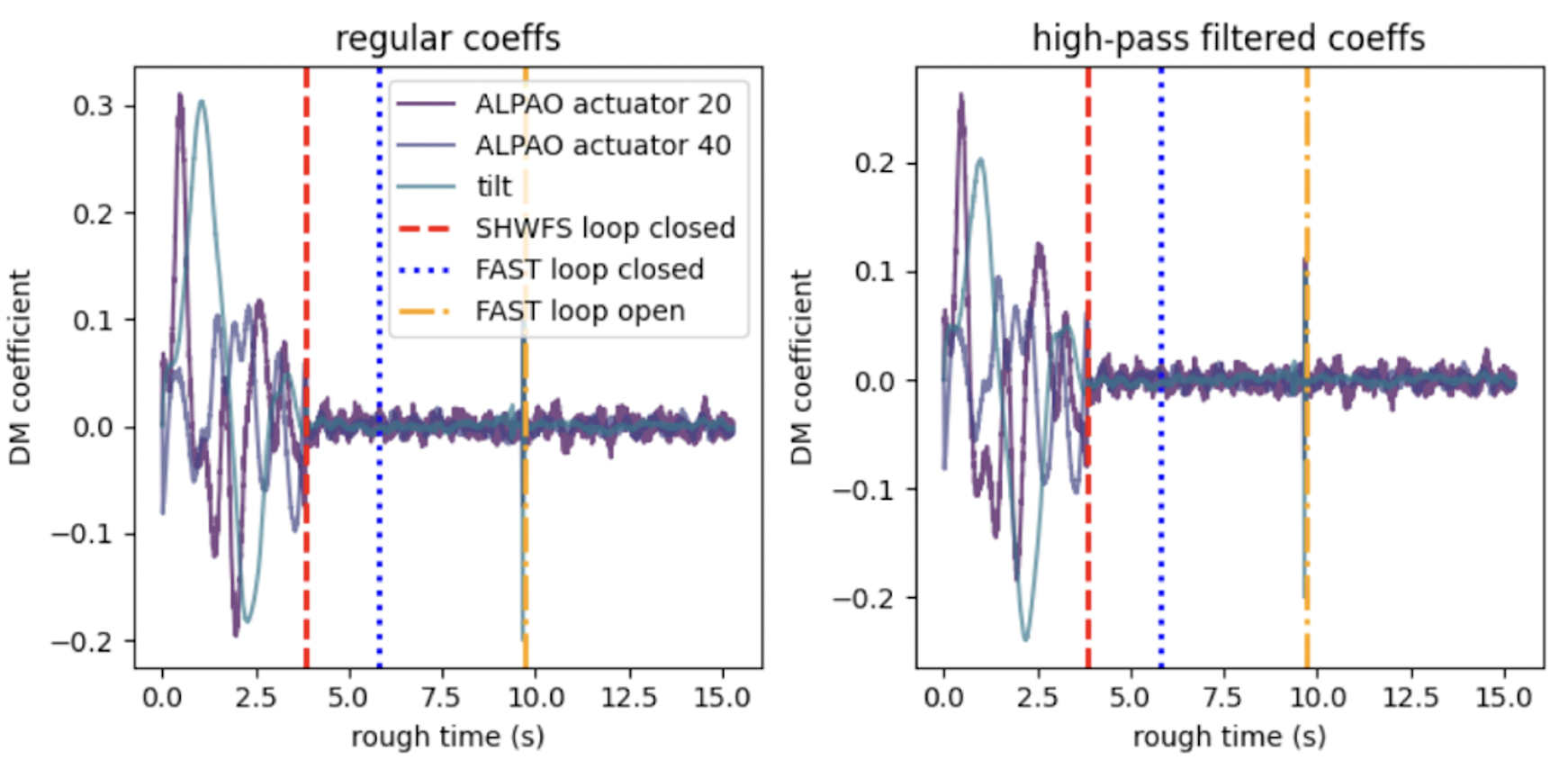}
        \caption{Non-zero FAST loop gain}
    \end{subfigure}    
    \vspace*{5pt}
    \caption{SHWFS telemetry for a FAST + SHWFS Multi-WFS SCAO laboratory demonstration. In each panel, DM-injected disturbances over the full sequence simulate open loop atmospheric wavefront error, normalized to 1 $\mu$m rms (although the y axis is shown in DM units that are not normalized to physical units). At the dashed red line, the SHWFS loop closes with full control authority over all temporal frequencies. At the dotted blue line, the FAST loop closes, at which point the FAST and SHWFS loops switch to using low- and high-pass filtered coefficients, respectively, as described in \S\ref{sec: concept_and_setup}. At the dash dotted orange line, the FAST loop opens again and the SHWFS loop switches back to full control authority of all temporal frequencies. In panel (a) the FAST loop gains are set to zero (i.e., triggering a switch in closed-loop to temporally filtered coefficients but effectively opening the FAST loop), while the gains are non-zero in panel (b). Unfiltered and high-pass filtered SHWFS coefficients are shown in the left and right sub-panel of each panel, respecively.}
    \label{fig: results}
\end{figure}

Figure \ref{fig: results} shows the results of multi-WFS SCAO with a SHWFS and FAST closing the loop and keeping the loop closed.
Fig. \ref{fig: results} can be interpreted as follows:
\begin{enumerate}
    \item In panel a, compare the regular and high-pass filtered SHWFS coefficients in between the dotted blue and dash dotted orange lines (i.e., where the FAST and SHWFS loops switch to temporally filtered coefficients but with the FAST loop gain set to zero), the high-pass filtered coefficients show no difference from when the SHWFS has full control authority. This is because the SHWFS loop runs at faster speeds than FAST, and so the errors produced by not closing the loop with FAST are set to be at lower temporal frequencies than are seen by this high-pass filter.
    \item Then, in panel b, the same comparison clearly shows in the unfiltered coefficients that the FAST loop remains closed and stable at a similar level of residuals as when the SHWFS loop is closed but with full control authority. The closed-loop region to the right of the dashed red line is essentially identical between unfiltered and filtered coefficients (left and right panel), showing a successful laboratory demonstration of this multi-WFS SCAO technique.
\end{enumerate}
\section{CONCLUSIONS AND FUTURE WORK}
\label{sec: conclusions}
In this paper we built on the multi-WFS SCAO framework, presented first in Ref. \citenum{gerard21a}, showing the first laboratory demonstration of the concept. The concept relies on using spatial and temporal filtering to allow two or more WFSs to point at the same star and close the loop with one or more common path DMs. In this laboratory demonstration on UCSC's SEAL testbed, we have used a 23$\times$23 SHWFS operating at 200 Hz and FAST operating at 100 Hz to both control low order modes with a 97-actuator ALPAO DM. In \S\ref{sec: results} We successfuly demonstrated this multi-WFS SCAO concept for both a loop closing acquisition sequence as well as steady-state operations in a stable configuration.

Future work can further mature the readiness of this technique to be applied on-sky. First and foremost, a laboratory demonstration of high+low order control will be critical for high contrast imaging applications; FAST dark hole images were not shown here because low order control does not produce dark holes, and so FAST images were essentially unchanged with and without FAST loops closed. However, we expect multi-WFS SCAO with low+high order control to show a clear benefit with the FAST loops closed, suggested by simulations in Ref. \citenum{gerard21a}. Implementing this software into a real-time low-level real-time control software may also be beneficial in future laboratory demonstrations, particularly in exploring performance gains over different temporal filter cutoff frequencies with minimal jitter between the timing of multiple loops at multiple frame rates, as discussed in \S\ref{sec: concept_and_setup}. Lastly, applying multi-WFS SCAO to other pairs of WFSs would be particularly interesting in laboratory demonstrations to compare with analagous cascaded AO configurations; in principle multi-WFS SCAO has simpler hardware requirements in not requiring a non-common path DM, but the added software complexity and related stability issues may be less ideal, and so this should be explored further.
\acknowledgements
This work was performed under the auspices of the U.S. Department of Energy by Lawrence Livermore National Laboratory under Contract DE-AC52-07NA27344. This document number is LLNL-PROC-852937.

\bibliography{report} 
\bibliographystyle{spiebib} 

\end{document}